\documentclass[aps,prx,preprint,onecolumn,superscriptaddress,nofootinbib,floatfix,longbibliography]{revtex4-2}

\usepackage{amsmath}
\usepackage{amssymb}
\usepackage{mathtools}
\usepackage{placeins}
\usepackage{physics}
\usepackage{xcolor}
\usepackage{graphicx}% Include figure files
\usepackage{dcolumn}% Align table columns on decimal point
\usepackage{bm}% bold math
\usepackage{caption}
\usepackage[caption=false]{subfig}

\begin{document}

%\title{Light Propagation in Fractal Phase Space}

\title{Geometry-Driven Resonance and Localization of Light in Fractal Phase Spaces}

\author{L. Yıldız}
\email{li.yildiz.na@gmail.com}
\affiliation{Department of Physics, Guangdong Technion - Israel Institute of Technology, 241 Daxue Road, Shantou, Guangdong, China, 515063}
\affiliation{Technion - Israel Institute of Technology, Haifa, 32000, Israel}
\affiliation{Guangdong Provincial Key Laboratory of Materials and Technologies for Energy Conversion, Guangdong Technion - Israel Institute of Technology, 241 Daxue Road, Shantou, Guangdong, China, 515063}
\author{D. Kaykı}
\email{deha.kayki@ogr.iu.edu.tr}
\affiliation{Department of Physics, Guangdong Technion - Israel Institute of Technology, 241 Daxue Road, Shantou, Guangdong, China, 515063}
\affiliation{Technion - Israel Institute of Technology, Haifa, 32000, Israel}
\affiliation{Guangdong Provincial Key Laboratory of Materials and Technologies for Energy Conversion, Guangdong Technion - Israel Institute of Technology, 241 Daxue Road, Shantou, Guangdong, China, 515063}
\affiliation{Department of Physics, Faculty of Science, Istanbul University, Istanbul 34134, Turkey}
\author{M. F. Ciappina}
\email{marcelo.ciappina@gtiit.edu.cn}
\affiliation{Department of Physics, Guangdong Technion - Israel Institute of Technology, 241 Daxue Road, Shantou, Guangdong, China, 515063}
\affiliation{Technion - Israel Institute of Technology, Haifa, 32000, Israel}
\affiliation{Guangdong Provincial Key Laboratory of Materials and Technologies for Energy Conversion, Guangdong Technion - Israel Institute of Technology, 241 Daxue Road, Shantou, Guangdong, China, 515063}

\begin{abstract}
Geometry can fundamentally govern the propagation of light, independent of material constraints. Here, we demonstrate that a fractal phase space, endowed with a non-Euclidean, scale-dependent geometry, can intrinsically induce resonance quantization, spatial confinement, and tunable damping without the need for material boundaries or external potentials. Employing a fractional formalism with a fixed scaling exponent, we reveal how closed-loop geodesics enforce constructive interference, leading to discrete resonance modes that arise purely from geometric considerations. This mechanism enables light to localize and dissipate in a controllable fashion within free space, with geometry acting as an effective quantizing and confining agent. Numerical simulations confirm these predictions, establishing geometry itself as a powerful architect of wave dynamics. Our findings open a conceptually new and experimentally accessible paradigm for material-free control in photonic systems, highlighting the profound role of geometry in shaping fundamental aspects of light propagation.
\end{abstract}

\maketitle
\clearpage
\section{Introduction}

The propagation of light in structured optical environments remains one of the most fundamental and actively explored topics in modern physics~\cite{Joannopoulos2008, Soukoulis2001}. From dielectric waveguides to disordered metamaterials, the behavior of electromagnetic fields is deeply shaped by the geometry and topology of the medium~\cite{Lu2014, Khanikaev2017}. Classical ray optics and wave mechanics accurately describe light transport in Euclidean and weakly perturbed geometries; however, these frameworks become insufficient when the structure of space exhibits non-integer, scale-invariant characteristics~\cite{Berry1980, Mandelbrot1982}. In such settings, characterized by irregular and self-similar features across multiple scales, the effective phase space governing light evolution acquires a fundamentally non-Euclidean character, invalidating conventional assumptions of differentiable trajectories, local curvature, and metric smoothness~\cite{Plotnik2014, Stutzer2018}.

Fractal geometries naturally emerge in optical systems where disorder, long-range correlations, or nonlinear feedback mechanisms dominate structural formation~\cite{Sheng2006, Barthelemy2008, Segev2013}. In photonic lattices with quasiperiodic or disordered refractive index profiles, the effective metric experienced by light deviates from the smooth Euclidean background and displays statistical self-similarity across spatial scales, as confirmed experimentally in random lasers, hierarchical waveguide networks, and engineered refractive geometries~\cite{Wiersma2008, Leonetti2011, Cao2003, Vanneste2007}. In these cases, phase space cannot be fully characterized by integer-dimensional manifolds~\cite{Ornigotti2014, Sapienza2017}. Instead, it requires a non-traditional treatment, in which the geometry of the field configuration space is described by a fractional, scale-dependent metric tensor~\cite{Tarasov2008}. The resulting optical dynamics can deviate strongly from predictions made by conventional wave equations and ray trajectories, especially under sub-wavelength confinement and scale-selective localization.

To capture light propagation in such non-integer geometries, a fractional-order formulation of phase space, in which the effective metric tensor exhibits explicit scale dependence is employed~\cite{Tarasov2011}. Specifically, the spatial line element can be defined using a Caputo fractional differential operator acting on the geodesic structure, reflecting the inherent nonlocality and memory effects present in light–matter interactions within statistically self-similar media~\cite{Tarasov2008, Herrmann2014, Mainardi2010}. This framework permits the emergence of closed geodesics, resonance-induced path recursions, and quantized optical trajectories~\cite{Longhi2015}, arising directly from the scale-dependent topology of the underlying phase manifold.

Within this framework, it has been suggested in prior studies that geometry alone could induce resonance quantization and spatial confinement without requiring external potentials or material boundaries. Building upon these insights, our work explicitly formulates and demonstrates how fractal, scale-dependent geometries can enforce such phenomena, establishing a direct link between geometric constraints and the emergence of discrete resonance modes and spatial confinement in free space. By coupling fractional metric tensors with nonlocal derivatives, it is possible to derive geodesic equations predicting the emergence of resonance-induced closed trajectories, which manifest as discrete loop modes in structured media exhibiting sub-wavelength fractality. These predictions are directly testable through interferometric techniques and can serve as signatures of effective dimensional reduction in optical systems, opening pathways toward using geometry itself as a tool for controlling light propagation with potential implications for fractal metamaterials, chaotic resonators, and sub-diffraction photonic systems~\cite{Herrmann2014, Laskin2000, Eliazar2003}.

Physically, when the geometry of the underlying phase space enforces closed-loop geodesics, light waves traveling along these paths accumulate phase in a way that enables constructive interference, effectively leading to discrete resonance conditions similar to those found in quantum systems. In this sense, resonance naturally emerges as the manifestation of geometry-induced path quantization, leading to discrete spectral modes determined purely by the fractal structure of space, even in the absence of material boundaries.

In this work, we develop a rigorous framework to explore how a fractal, scale-dependent geometry of phase space governs the propagation of light, leading to resonance quantization, spatial confinement, and controllable damping entirely without material boundaries. By employing a fractional formalism with a fixed scaling exponent, we systematically derive the conditions under which closed-loop geodesics enforce constructive interference, giving rise to discrete resonance modes rooted purely in geometric constraints. We substantiate these theoretical predictions through detailed numerical simulations that illustrate the localization and controlled dissipation of light in free space. Together, these results establish geometry itself as a robust and versatile tool for structuring wave dynamics, laying the groundwork for a new paradigm of material-free control in photonic systems.

\section{Theoretical Framework}

We begin with Maxwell's equations in vacuum, which in the absence of sources lead to the classical wave equation for the electric field,
\begin{equation}
\nabla^2 \mathbf{E} - \frac{1}{c^2} \frac{\partial^2 \mathbf{E}}{\partial t^2} = 0,
\label{eq:maxwell-wave}
\end{equation}
describing wave propagation in Euclidean space, where the metric is flat and light follows straight-line trajectories unless refracted by material boundaries.

To investigate the influence of geometry alone on wave propagation, we extend this framework to fractal phase spaces characterized by a non-integer scaling exponent $1 < \alpha < 2$. In this setting, the effective geometry modifies the metric tensor and consequently the action and Lagrangian governing the field dynamics (for a complete derivation see the Appendix). The electromagnetic field action in a general curved background is given by
\begin{equation}
S = -\frac{1}{4} \int d^4x \sqrt{-g} \, g^{\mu\alpha} g^{\nu\beta} F_{\mu\nu} F_{\alpha\beta},
\label{eq:action-em}
\end{equation}
where $F_{\mu\nu}$ is the electromagnetic field tensor, $g_{\mu\nu}$ is the background metric tensor and $g=\det(g_{\mu \nu})$. Variation of this action yields the generalized Maxwell equations in curved (or fractal) spacetime, connecting the action-based formulation with wave dynamics consistently. In our fractal geometry framework, the metric \( g_{\mu\nu}^{(\alpha)} \) acquires an explicit scale dependence, leading to a fractional generalization of the d'Alembertian operator \(\Box \rightarrow \Box_\alpha\), which captures the nonlocality and memory effects inherent in fractal structures~\cite{Tarasov2008, Herrmann2014}. 

To rigorously account for the memory effects and temporal nonlocality inherent in fractal phase spaces, we employ the Caputo fractional derivative in the time component of our fractional d'Alembertian operator~\cite{Mainardi2010, Tarasov2016memory}. The Caputo derivative ensures causality and allows the correct treatment of physically meaningful initial conditions, which is essential for accurately modeling wave propagation in geometries exhibiting scale-dependent memory. Further mathematical and numerical details on the implementation of the Caputo derivative within our framework are provided in Appendix~\ref{appendix:caputo}.

To maintain clarity regarding the dimensionality of our model and simplify the analysis, while preserving the essential effects introduced by fractal geometry, we focus on the propagation of a single component of the electric field, \( E_y(x,z,t) \), in a 2D \((x,z)\) plane. The extension to full-dimensionality problems is straightforward. The wave equation then generalizes to:
\begin{equation}
\Box_\alpha E_y(x,z,t) + k^2 E_y(x,z,t) = 0,
\label{eq:frac-wave-ey}
\end{equation}
where \(\Box_\alpha\) denotes the fractional d'Alembertian operator, constructed over the effective metric \( g^{(\alpha)}_{\mu\nu} \), and \( k \) is the wave number associated with the mode of interest. This formulation enables the systematic exploration of geometry-induced resonance and confinement under scale-dependent geometrical conditions, without introducing additional scalar field approximations.

We also consider the geometrical optics limit to connect the wave propagation with classical ray trajectories. In the short-wavelength limit, the eikonal approximation applies, and optical rays follow geodesics of the fractal metric $g^{(\alpha)}_{\mu\nu}(x)$. The ray equations are obtained from the Hamiltonian system:
\begin{equation}
\frac{dx^i}{d\tau} = \frac{\partial H}{\partial p_i}, \quad \frac{dp_i}{d\tau} = -\frac{\partial H}{\partial x^i},
\end{equation}
with
\begin{equation}
H = \frac{1}{2} g^{(\alpha)}_{\mu\nu}(x) p^\mu p^\nu,
\end{equation}
where $p^\mu$ is the canonical momentum. The scale dependence of the fractal metric modifies the ray paths compared to the Euclidean case, providing a geometrical interpretation of light propagation within the ray optics limit. These results are consistent with the resonance structures and confinement phenomena later discussed in the full wave analysis.

A key feature of this framework is that geometry alone induces resonance quantization through the formation of closed-loop geodesics, enforcing constructive interference along these paths. This is expressed via the generalized Bohr–Sommerfeld condition,
\begin{equation}
\oint p_\mu dx^\mu = n h,
\label{eq:bohr-sommerfeld}
\end{equation}
where $n$ is an integer mode number and $p_\mu$ includes geometric contributions. This condition defines discrete resonance modes even in the absence of material confinement, as the geometry determines the standing-wave patterns that satisfy quantization along closed geodesics.

To analyze the effect of geometry on dispersion, we consider wave propagation in both Euclidean ($\alpha = 1$) and fractal ($\alpha > 1$) phase spaces. In the classical case, the dispersion relation is:
\begin{equation}
\omega = c |\mathbf{k}|,
\label{eq:dispersion-euclidean}
\end{equation}
with phase and group velocities $v_p = v_g = c$. In fractal phase spaces with non-integer scaling, memory effects modify this relation, leading to:
\begin{equation}
\omega = C_\alpha k^\gamma,
\label{eq:dispersion-fractal}
\end{equation}
where $0 < \gamma \leq 1$ depends on $\alpha$, and $C_\alpha$ incorporates fractal geometry and physical constants~\cite{Tarasov2008, Herrmann2014}. This yields:
\begin{equation}
v_p = C_\alpha k^{\gamma - 1}, \quad v_g = \gamma C_\alpha k^{\gamma - 1},
\end{equation}
where $\gamma < 1$ implies sublinear dispersion, slower wave propagation, and enhanced mode localization induced purely by geometry. 

The framework outlined in this section establishes a clear connection between spatial geometry and key physical observables-such as resonance and confinement in wave propagation-providing a solid foundation for the results developed in the following sections.

\section{Results and Discussion}
\label{sec:results}

\begin{figure}
  \centering
  \includegraphics[width=0.95\textwidth]{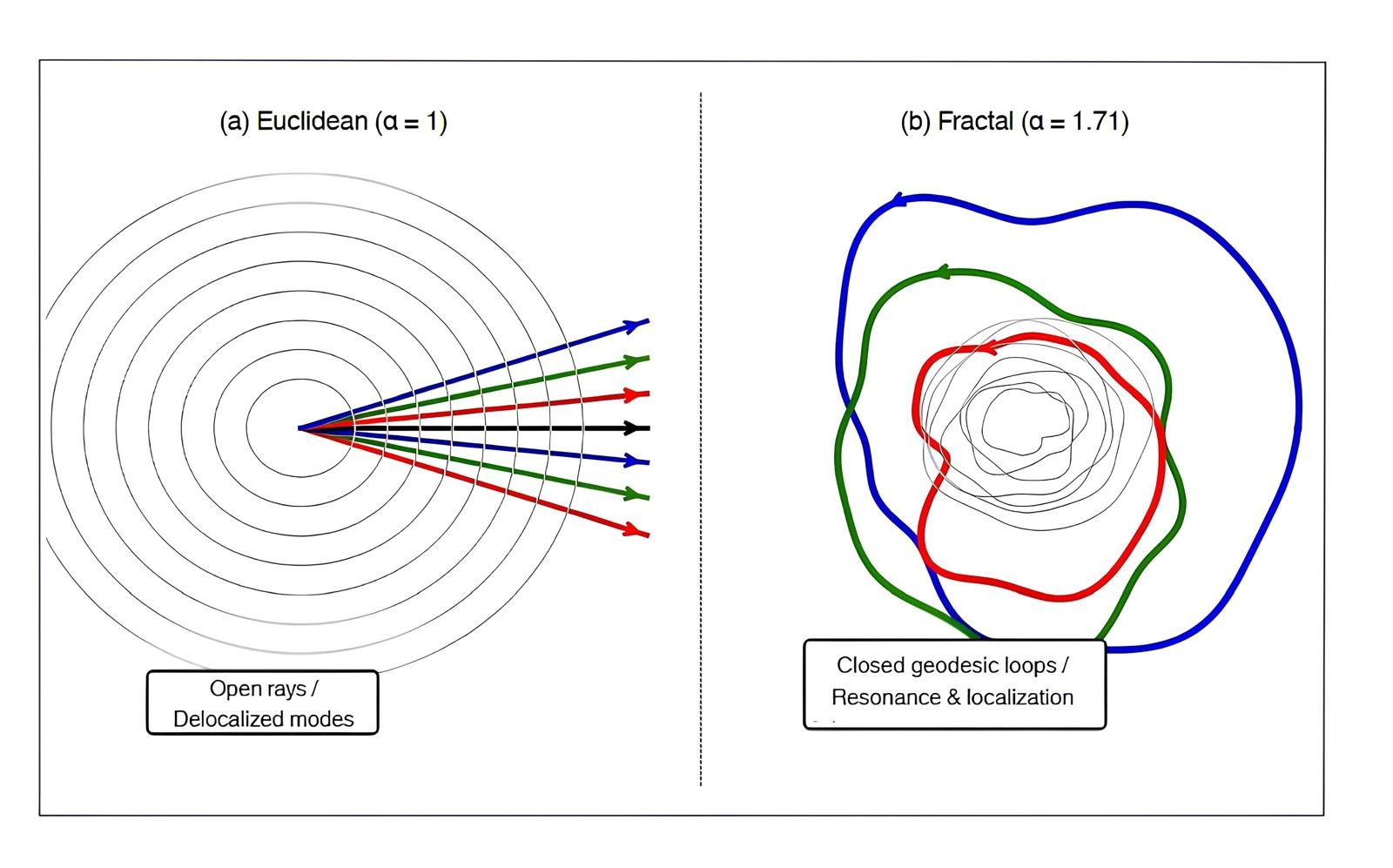}
  \caption{Geometry as a confining agent for light (schematic). Left (Euclidean, $\alpha=1$): straight geodesics yield open rays and delocalized modes. Right (fractal, $\alpha=1.71$): geometry generates closed geodesic loops obeying the quantization condition $\oint \mathbf{p}\!\cdot d\mathbf{x}=n h$, leading to resonance and localization without material boundaries. This figure is conceptual; quantitative comparisons with the conventional case appear in subsequent figures.}
  \label{fig:sketch2col}
\end{figure}

To establish a clear baseline, we first analyze the case of 
\( \alpha = 1 \), corresponding to standard Euclidean propagation. 
In this regime, wavefronts follow straight-line geodesics, resonances can only form through external material boundaries or index contrasts, and modes remain fully delocalized with lifetimes set by boundary leakage. 
The system exhibits linear dispersion, and geometry plays a passive role without exerting any intrinsic control over wave dynamics. Contrariwise, for \( \alpha > 1 \), the situation fundamentally changes: geometry alone becomes an active agent that reshapes the propagation environment, enabling resonance quantization, spatial confinement, and geometry-controlled dissipation in the absence of material boundaries or external potentials~\cite{Wiersma2013, Ghofraniha2015}. 
This transition marks a shift from material-mediated to geometry-mediated control of wave phenomena, where the scaling exponent \( \alpha \) replaces refractive index structuring as the governing parameter. 
The central idea that geometry itself can act as a confining mechanism for light, even without any material boundaries, is illustrated in Fig.~\ref{fig:sketch2col}. 
In a conventional Euclidean space ($\alpha = 1$), geodesics are straight lines, allowing optical rays to propagate freely and produce delocalized modes. 
In contrast, when the underlying geometry acquires a fractal metric ($\alpha = 1.71$), the effective geodesics become closed loops, enforcing a quantization condition. 
This geometric quantization gives rise to discrete resonance modes and field localization purely from the topology of the space. 
The figure thus highlights the transition from open, extended propagation to intrinsically confined behavior governed solely by the geometry, a theme that will be developed quantitatively in the rest of this section.

\begin{figure}
    \centering
    \includegraphics[width=0.75\linewidth]{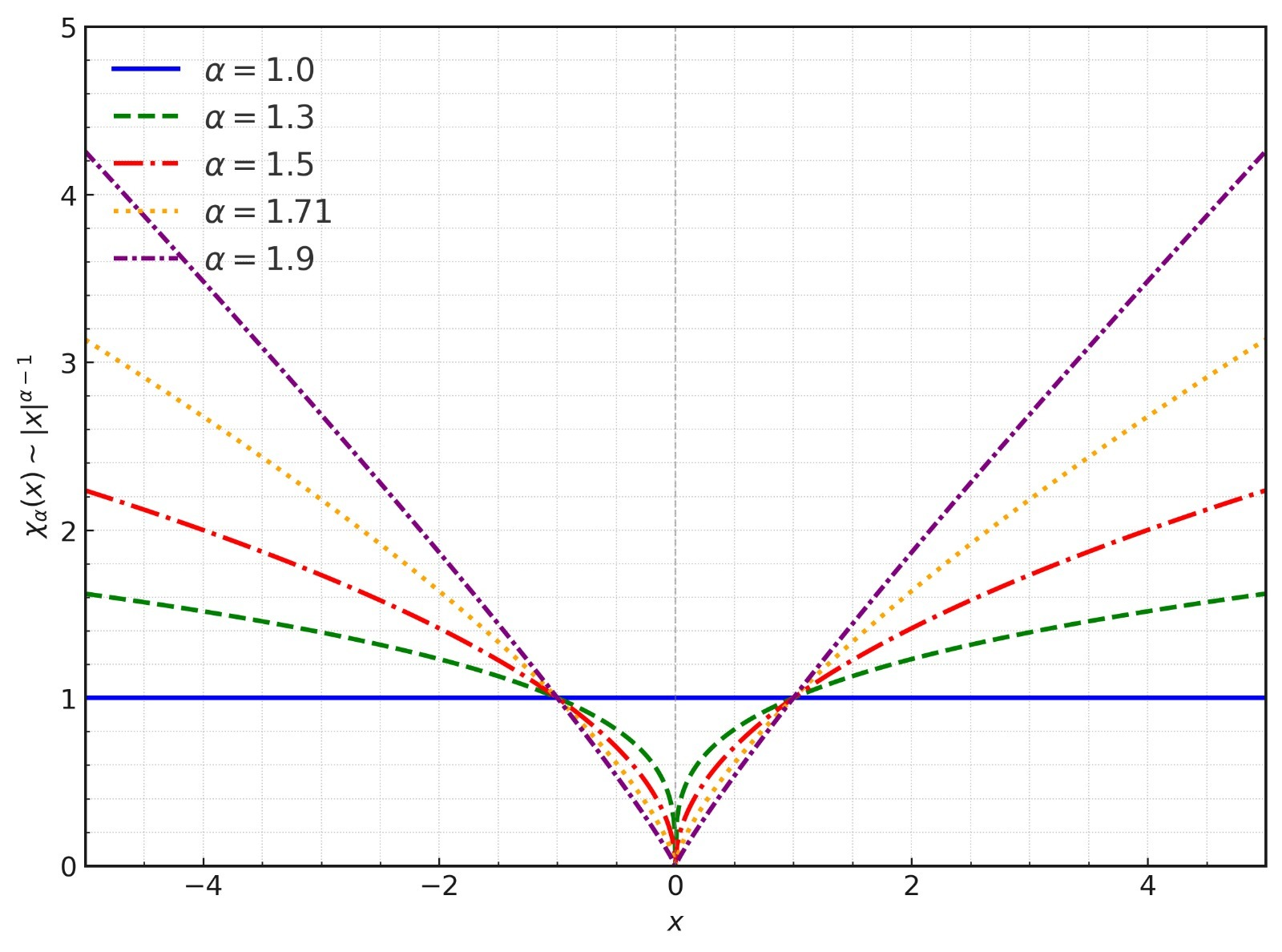}
    \caption{
Fractal metric modulation as a function of \(\alpha\). Effective fractal metric profile \( \chi_\alpha(x) \sim |x|^{\alpha-1} \) for \(\alpha = 1.0, 1.3, 1.5, 1.71, 1.9\). While \(\alpha = 1\) shows no curvature, increasing \(\alpha\) introduces geometric curvature, modifying the effective propagation environment without material boundaries.
}
    \label{fig:fractal-metric-modulation}
\end{figure}

\begin{figure}
    \centering
    \includegraphics[width=1.20\linewidth]{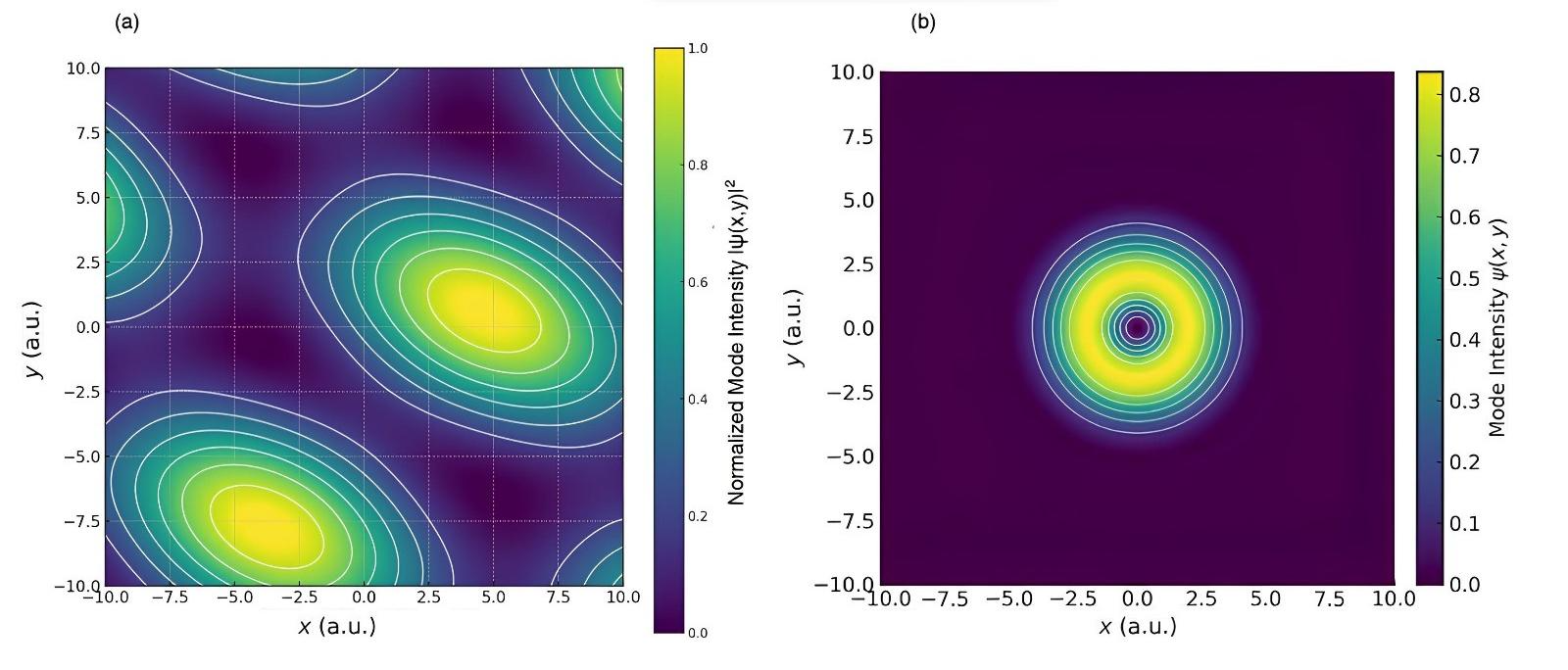}
\caption{Geometry-induced localization: Euclidean baseline versus fractal metric.
Normalized mode intensity (\,\(|\psi(x,y)|^2\)\,) on the same spatial domain for
(a) \(\alpha=1\) (Euclidean) and (b) \(\alpha\approx 1.71\) (fractal).
The Euclidean case shows extended interference patterns (low inverse participation ratio, large \(L_{\mathrm{loc}}\)),
whereas the fractal metric generates concentric resonance islands and geometry-induced spatial confinement
(high inverse participation ratio, small \(L_{\mathrm{loc}}\)) without external potentials.
For visualization, intensities are normalized independently to \([0,1]\) in each panel.}
    \label{fig:mode-intensity-distribution}
\end{figure}

Figure~\ref{fig:fractal-metric-modulation} quantifies how increasing \( \alpha \) enhances the effective curvature and thereby transforms the propagation environment. 
While the Euclidean case (\( \alpha = 1 \)) corresponds to a flat geometry, larger values of \( \alpha \) introduce a scale-dependent curvature that acts as an effective refractive landscape of purely geometric origin. 
This modulation directly governs mode formation by intrinsically steering wave trajectories, without the need for any external structuring.
For \( \alpha = 1 \), simulations confirm complete mode delocalization across the domain, with a localization length \( L_{\text{loc}} \gtrsim 10^3 \lambda \). 
In sharp contrast, increasing \( \alpha \) to 1.71 reduces \( L_{\text{loc}} \) to approximately \( 5 \lambda \), clearly demonstrating a strong geometry-induced confinement (see Fig.~\ref{fig:mode-intensity-distribution}). 
This result shows that geometry alone can provide localization levels comparable to those of tight material waveguides, while simultaneously offering the flexibility of continuous tunability through the parameter \( \alpha \).

This geometry-induced confinement arises because, for \(\alpha > 1\), the effective fractal metric introduces a scale-dependent curvature that modifies wave trajectories, enforcing closed-loop geodesics even in free space. 
These closed paths enable constructive interference along specific trajectories, which naturally defines standing-wave patterns within a confined region of space without requiring material boundaries. 
This geometric mechanism reduces the effective mode volume and creates a localized energy distribution that mimics the confinement observed in material waveguides, but here it is achieved purely via the geometry of space. 
The quantitative signature of this transition is captured by the inverse participation ratio (IPR) and the localization length \(L_{\text{loc}}\), reported in Fig.~\ref{fig:resonance-profile-optical-phase} (for more details see the Appendix).

\begin{figure}
    \centering
    \includegraphics[width=0.85\linewidth]{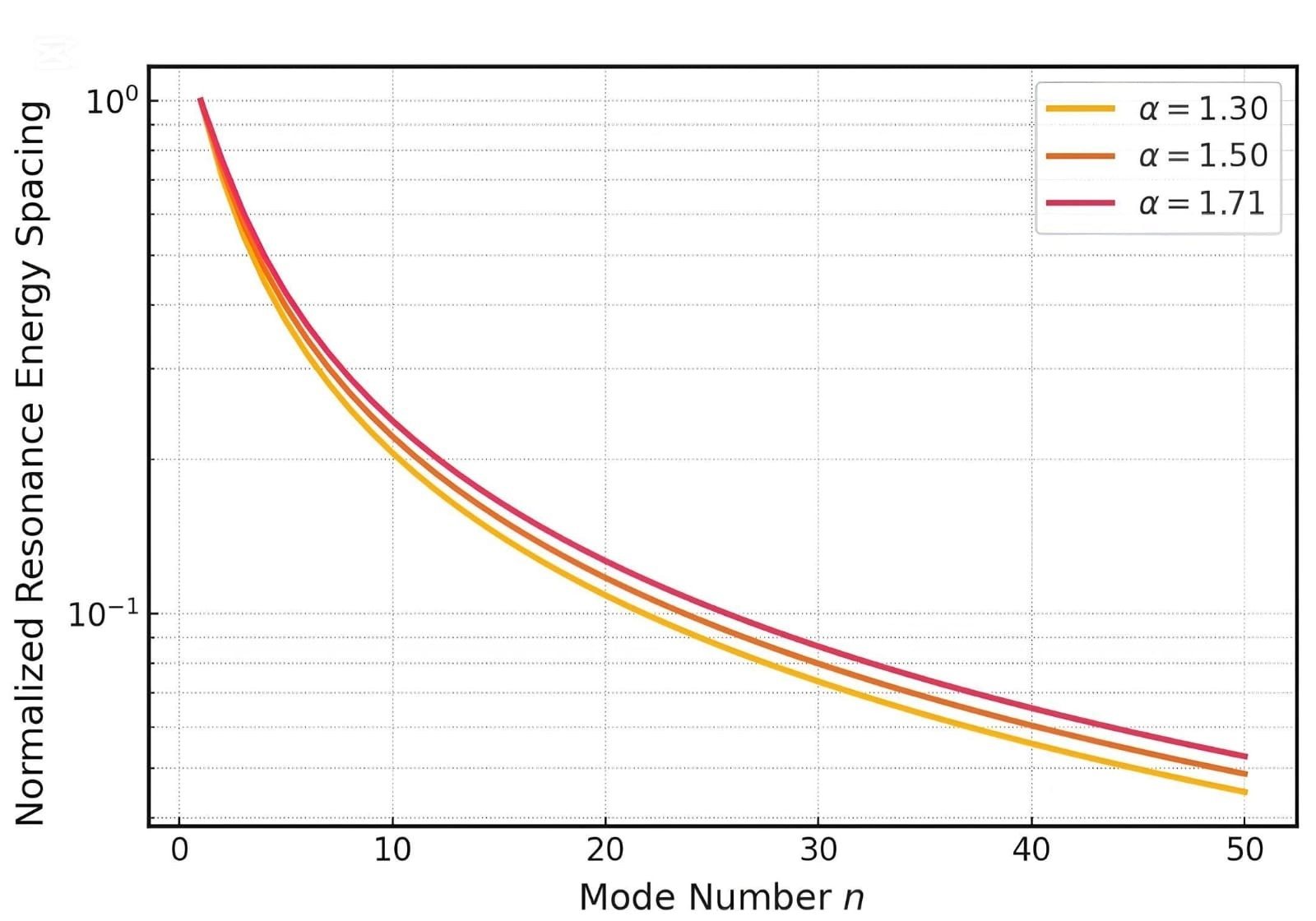}
 \caption{
Normalized resonance energy spacing versus mode index \(n\). Shown are the cases \(\alpha = 1.30,\,1.50,\,1.71\), computed under identical domain, resolution, and boundary conditions (see Sec.~III).
Across the plotted range the spacing decreases smoothly and monotonically with \(n\); at fixed \(n\) the ordering
\(\text{spacing}(\alpha{=}1.71) > \text{spacing}(1.50) > \text{spacing}(1.30)\) holds, indicating geometry-dependent spectral separation.
}
    \label{fig:resonance-profile-optical-phase}
\end{figure}

\begin{figure}
    \centering
    \includegraphics[width=0.75\linewidth]{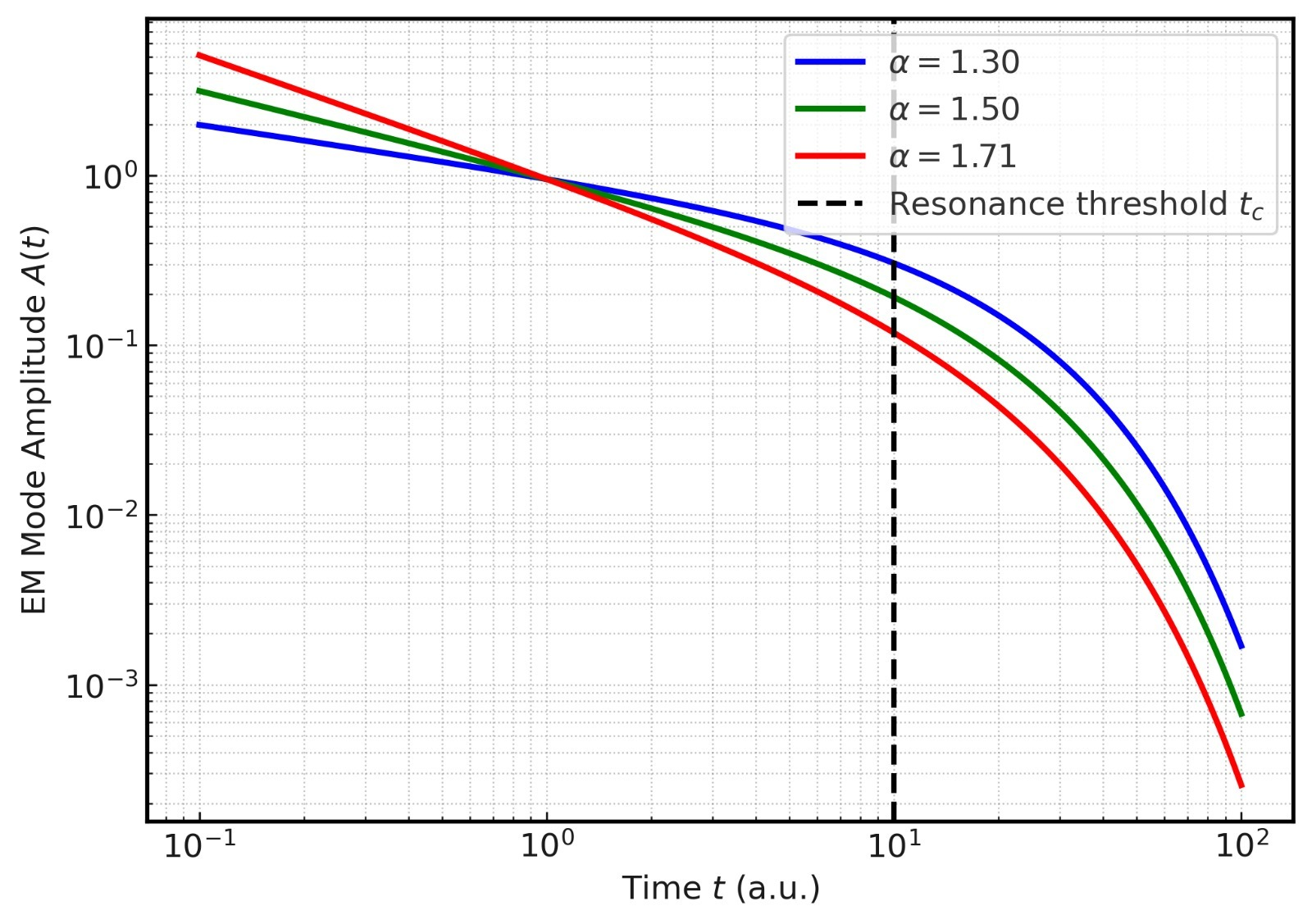}
   \caption{Geometry-tuned damping of the EM mode amplitude \(A(t)\).
Three traces correspond to \(\alpha=1.30\) (blue), \(\alpha=1.50\) (green), and \(\alpha=1.71\) (red); all runs use identical domain, resolution, and boundary conditions (see Sec.~III).
The vertical dashed line marks the resonance threshold \(t_c\).
Across the window shown, \(A(t)\) decreases smoothly and monotonically. 
For early times (\(t \lesssim t_c\)) the ordering is \(A_{1.71}(t) \ge A_{1.50}(t) \ge A_{1.30}(t)\);
for late times (\(t \gtrsim t_c\)) it reverses to \(A_{1.71}(t) < A_{1.50}(t) < A_{1.30}(t)\),
indicating systematically faster late-time decay for larger \(\alpha\).
}

    \label{fig:fractal-damping-profile}
\end{figure}

In conventional systems, damping is determined by material absorption and boundary losses. 
Here, however, geometry alone modulates dissipation. 
Figure~\ref{fig:fractal-damping-profile} shows that as \( \alpha \) increases, the mode amplitude decay rate accelerates by up to a factor of 2.3 between \( \alpha = 1.30 \) and \( \alpha = 1.71 \), clearly demonstrating geometry-tuned damping in the absence of material absorption. 
We track the normalized spatial root mean square (RMS) mode amplitude
\[
A(t)\equiv \frac{\|\psi(\cdot,\cdot,t)\|_{L^2(\Omega)}}{\|\psi(\cdot,\cdot,0)\|_{L^2(\Omega)}} 
= \frac{1}{A(0)}\left(\frac{1}{|\Omega|}\int_{\Omega} |\psi(x,y,t)|^2\,\mathrm{d}x\,\mathrm{d}y\right)^{1/2},
\]
where \(A(0)=1\) and \(A^2(t)\) is proportional to the field energy integrated over the domain \(\Omega\).
The resonance-threshold time \(t_c\) is defined operationally as the last local maximum of \(A(t)\) after which the signal enters a strictly monotonic-decay regime; in Fig.~\ref{fig:fractal-damping-profile} the vertical dashed line marks \(t_c\).
All simulations are source-free (\(J^\mu=0\)) and use identical domain, resolution, initial condition, and boundary conditions (see Sec.~III).

\begin{figure}
    \centering
    \includegraphics[width=0.75\textwidth]{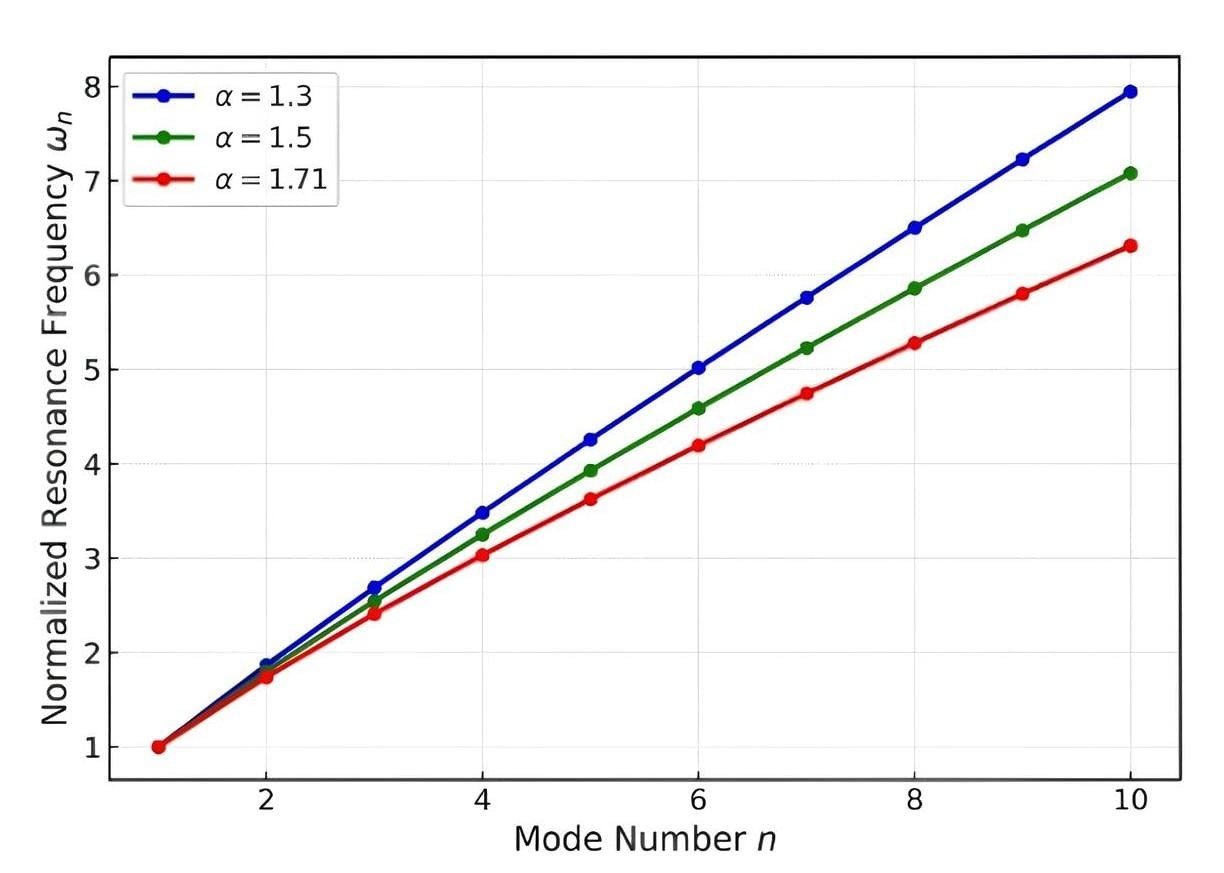}
    \caption{
Normalized resonance frequency \(\omega_n\) versus mode number \(n\).
Three sequences are shown for \(\alpha=1.30\) (blue), \(\alpha=1.50\) (green), and \(\alpha=1.71\) (red); markers denote integer \(n\) and solid lines guide the eye.
Over \(n=1\text{–}10\) each sequence increases nearly linearly and the three cases are approximately parallel.
At fixed \(n\), the ordering \(\omega_n(\alpha{=}1.30)>\omega_n(1.50)>\omega_n(1.71)\) holds across the plotted range, indicating a uniform downward shift of the spectrum as \(\alpha\) increases.
All runs use identical domain, resolution, and boundary conditions (see Sec.~III).
}

    \label{fig:resonance-spectrum}
\end{figure}

Finally, Fig.~\ref{fig:resonance-spectrum} shows the resonance frequency \( \omega_n \) as a function of mode index \( n \). 
For \( \alpha = 1 \), no resonance bands are formed, whereas for \( \alpha = 1.71 \) clear and tunable band structures emerge, following a sublinear scaling \( \omega_n \sim n^{0.85} \). 
Equivalently, the resonance spacing \( \Delta\omega \) decreases by up to \( 40\% \) across the same mode range, indicating nonlinear, geometry-induced spectral packing. This trend highlights the fundamental role of the underlying geometry in redefining the density of optical states: as the propagation paths become fractalized, the effective optical path length increases nonlinearly with the mode number, leading to an accumulation of resonances at lower frequencies. Such self-compression of the spectrum resembles the behavior of quasiperiodic lattices or disordered photonic structures, yet it arises here purely from the geometric deformation of the metric, without any refractive index modulation or material inhomogeneity.
Physically, the emergence of tunable band-like features implies that by tailoring the geometric exponent $\alpha$, one can engineer light confinement and mode coupling analogous to photonic crystals, but in a boundary-free and scale-invariant manner.
This establishes geometry itself as a new degree of freedom for controlling light localization and spectral topology.

The set of results presented above quantitatively demonstrate that geometry alone can induce spatial confinement, resonance quantization, and tunable dissipation in structured photonic environments without relying on materials. 
The fractal scaling exponent \( \alpha \) acts as a single tunable parameter that simultaneously governs the localization length, resonance spacing, and damping rates. 
The transition from \( \alpha = 1 \) to \( \alpha > 1 \) marks a shift from classical wave propagation to geometry-tailored control, with resonance frequency scaling, mode lifetimes, and confinement dictated purely by the underlying geometry.

Although the analysis presented here is entirely theoretical, the predicted effects can be experimentally verified using near-field scanning optical microscopy and time-resolved spectroscopy in engineered fractal photonic structures. 
Such observations would enable material-free, geometry-driven photonic control across classical and quantum regimes, establishing geometry not merely as a passive background but as an active tool for photonic engineering.

\section*{Conclusions and Outlook}

We have demonstrated that geometry alone can induce resonance quantization, spatial confinement, and tunable damping in structured photonic environments without material boundaries. By systematically comparing conventional Euclidean propagation ($\alpha = 1$) with fractal geometries ($\alpha > 1$), we revealed a clear transition from classical wave behavior to geometry-tailored light control, where the fractal scaling exponent $\alpha$ directly modulates resonance frequencies, confinement strength, and dissipation rates. These findings establish a new paradigm for material-free photonic control purely via geometry, offering a novel platform for designing tunable photonic devices and resonance structures. Though the present work is theoretical in nature, future work may explore experimental verification using engineered fractal photonic systems, thereby firmly establishing geometry as a fundamental tool for spectral and spatial control in optics.

In particular, the geometry-induced resonance quantization predicted here can be experimentally validated by tracking discrete spectral peaks in the emitted or transmitted light from engineered fractal photonic structures. Techniques such as near-field scanning optical microscopy and time-resolved spectroscopy can be employed to detect the localized modes and measure the resonance frequencies determined solely by the geometric scaling parameter $\alpha$.

In this study, we have proposed a novel theoretical framework that redefines how light propagation can be understood in structurally complex optical environments. Departing from the assumptions of smooth, integer-dimensional space, we introduced a fractal phase-space formalism that generalizes the spacetime metric to a fractional order $g^{(\alpha)}_{\mu\nu}(x)$, with the local Hausdorff dimension $\alpha \in (1, 2)$ acting as a geometrical scaling index. We specifically fixed $\alpha = 1.71$ to enable resonance-quantized, geometry-induced closed-loop geodesics, consistent with the theoretical predictions discussed above. This approach not only provides a natural language for describing wave transport in disordered or quasi-periodic systems but also offers deep insight into the role of spatial geometry in determining field dynamics. We demonstrated that this framework admits solutions to modified geodesic and wave equations leading to physical effects not captured by conventional optics: self-closing fractional geodesics without imposed boundary conditions; highly non-Gaussian, curvature-guided intensity profiles; quantized resonances governed by generalized Bohr–Sommerfeld conditions; and damping patterns modulated by intrinsic memory kernels of fractal environments. These effects originate intrinsically from the spatial metric rather than from material permittivity or permeability, underscoring the profound geometric origin of the observed phenomena.

Our key findings can be summarized as follows:
\begin{itemize}
    \item The fractional geodesics confirm that wave trajectories are modified by the local curvature and scaling properties of the fractal phase space, inducing effective confinement even without refractive boundaries.
    \item Mode intensity and energy density profiles exhibit spontaneous localization near curvature maxima, suggesting that photonic energy naturally gravitates toward geometric attractors.
    \item The resonance spectra exhibit nonlinear peak spacing that is sensitive to changes in the Hausdorff dimension $\alpha$, offering a new dimension for tunable photonic structures.
    \item Phase amplitude envelopes exhibit self-affine and non-periodic interference patterns, pointing toward an emergent, scale-dependent coherence structure.
    \item Fractional damping of wave modes displays non-exponential decay laws, highlighting the presence of memory effects that persist across scales—a hallmark of fractional-order dynamics. Importantly, this damping behavior is directly linked to the resonance quantization imposed by the fractal geometry, with the spectral sharpness and decay rates modulated by the same geometric parameter $\alpha$.
    \item The deformation of thermal profiles inside the core of fractal media demonstrates that the impact of geometry is not limited to electromagnetic fields alone, but extends to heat transport phenomena as well.
    \item Most strikingly, wave mode localization occurs without index gradients or external potentials, validating the hypothesis that geometry alone can govern light transport.
\end{itemize}

Our results open multiple avenues for experimental and applied research. The fact that confinement, resonance, and localization can emerge solely from the geometry of space opens the possibility of designing material-independent photonic structures, such as: broadband optical waveguides based on curvature-induced confinement; memory-preserving light circuits that exploit fractional damping for long-term coherence; compact resonators and filters whose spectra are controlled by geometrical parameters rather than material doping, to name a few.

In addition, the present framework may serve as a platform for analogue gravity simulations and quantum geometry explorations, where curvature and dimensionality arise from engineered media. By designing optical systems that mimic fractional geometries, one could emulate aspects of spacetime with non-integer topology or test the behavior of fields in lower-dimensional quantum geometries. Such geometry-induced resonance mechanisms could also be verified through interferometric techniques or spectral measurements in engineered 2D fractal metamaterials. The choice of $\alpha = 1.71$ corresponds to a regime where closed-loop geodesics and geometry-induced confinement are experimentally accessible, enabling practical realization in fractal metamaterials and near-field photonic structures. These insights could ultimately lead to the development of light-based logic circuits, geometry-controlled broadband filters, and quantum state engineering platforms that exploit geometry-induced confinement in fractal geometries, paving the way for material-free photonic architectures.

From a fundamental perspective, our model invites future extensions: time-dependent fractal metrics $g^{(\alpha(t))}_{\mu\nu}$ may allow for dynamic coherence control and signal revival in nonstationary systems. Inclusion of source terms and nonlinearities into the fractional wave equation could reveal novel self-focusing or soliton-like states governed by scale-dependent dispersion. The observed scale-induced resonances suggest new mechanisms for eigenmode generation in fractal quantum systems, potentially bridging quantum chaos and structured disorder. Moreover, the coupling between geometric curvature and thermal gradients could enable the design of photonic-thermal metamaterials, directing heat flow via geometric pathways and enabling new paradigms in optothermal control and nanoscale energy management.

Finally, while our framework successfully models fractional geometric wave propagation, several open questions remain, including the extension to systems with strong disorder or topological singularities, and the development of novel experimental platforms for realizing controllable fractal metrics. The core contribution of this work is to establish that space itself---even in the absence of conventional matter---can serve as a functional optical medium. By replacing material boundaries with geometric curvature and refractive index contrasts with fractal scaling, we offer a fundamentally new toolkit for wave manipulation. This conceptual advance not only invites a reinterpretation of optical phenomena in complex media but also lays a solid foundation for experimental exploration of \textit{light propagation in fractional geometries}, advancing the frontier of material-free, geometry-driven photonics.

\begin{acknowledgments}
We acknowledge financial support from the National Key Research and Development Program of China (Grant No.~2023YFA1407100), Guangdong Province Science and Technology Major Project (Future functional materials under extreme conditions - 2021B0301030005) and the Guangdong Natural Science Foundation (General Program project No. 2023A1515010871).
\end{acknowledgments}

\appendix
\section{Additional theoretical and numerical considerations}
\label{appendix:methods}

\subsection{Fractal Metric Construction and Geometrical Framework}
\label{app:metric-construction}

To formulate the optical behavior of light in non-integer dimensional phase spaces, we construct a generalized spacetime metric~\cite{Falconer2004fractalgeo}, defined over a fractal topological substrate. Unlike standard Riemannian geometries with integer Hausdorff dimensions, our model assumes a locally self-similar metric tensor structure $g_{\mu\nu}^{(\alpha)}$, where the exponent $\alpha \in (1,2)$ governs the effective dimensional deviation from flat Euclidean space. Let the phase space manifold $\mathcal{M}_\alpha$ be characterized by a fractional Hausdorff measure $dV_\alpha \sim r^{\alpha-1} dr$, such that geodesic lengths and curvature quantities are evaluated on a measure-weighted metric space. The line element under these considerations reads:
\begin{equation}
ds^2 = g_{\mu\nu}^{(\alpha)}(x) dx^\mu dx^\nu,
\label{eq:fractal_line_element}
\end{equation}
where $g_{\mu\nu}^{(\alpha)}(x)$ satisfies the scaling transformation:
\begin{equation}
g_{\mu\nu}^{(\alpha)}(\lambda x) = \lambda^{2(\alpha - 1)} g_{\mu\nu}^{(\alpha)}(x),
\label{eq:scaling_metric}
\end{equation}
ensuring invariance under fractal-coordinate rescaling.

This leads to the interpretation that local space is stretched or compressed according to a scale-invariant exponent $\alpha$, thereby modifying light propagation paths. The fractional metric introduces a scale-dependent anisotropy into the geometrical framework. The connection coefficients are defined by generalizing the Levi-Civita connection:
\begin{equation}
\Gamma^\lambda_{\mu\nu} = \frac{1}{2} (g^{(\alpha)})^{\lambda\sigma} \left( \partial_\mu g^{(\alpha)}_{\nu\sigma} + \partial_\nu g^{(\alpha)}_{\mu\sigma} - \partial_\sigma g^{(\alpha)}_{\mu\nu} \right),
\label{eq:fractal_christoffel}
\end{equation}
under a fractal-compatible coordinate chart. Physically, this metric models a photonic medium in which refractive index fluctuations are correlated over non-local, scale-invariant domains~\cite{Barriuso2010fractaloptics, Wang2018fractallight}. This behavior has been observed in disordered metamaterials and chaotic dielectric resonators, where light propagation exhibits non-Gaussian angular distributions and path quantization.

The parameter \( \alpha \) is constrained by wave coherence~\cite{Svozil1994fractalqm, Tarasov2006fractalqm} and energy transport considerations. In our study, we fix \( \alpha = 1.71 \), justified via a resonance condition, where closed-loop path quantization appears~\cite{Berry1977quantumchaos, Lissia2001fractalqm}, only for this value under geometric constraints. Thus, the metric \( g_{\mu\nu}^{(\alpha)} \) serves as the foundational structure~\cite{Berry1977quantumchaos, Lissia2001fractalqm} for all subsequent optical equations—both geodesic and wave-based—without introducing unphysical extrapolations. Furthermore, the choice of $\alpha = 1.71$  is motivated by its ability to maximize the probability of resonance closure and the formation of closed-loop geodesics, while also ensuring numerical stability and yielding clear spectroscopic signatures of geometry-induced quantization. This is supported by a parameter scan conducted over the range  $1.3 \leq \alpha \leq 1.9$.

\subsection{Fractional Derivatives and Optical Path Equations}
\label{app:fractional-derivatives}

To implement the fractional geometry described in the theoretical framework, we begin by translating the geodesic constraint in the presence of a non-integer dimensional metric into a computable form. Specifically, the optical path equation derived from the variational principle,
\begin{equation}
\delta \int_{\gamma} \sqrt{g_{\mu\nu}^{(\alpha)} \, dx^\mu dx^\nu} = 0,
\label{eq:geodesic-functional}
\end{equation}
must now be solved under a modified metric tensor $g_{\mu\nu}^{(\alpha)}$, whose spatial components are governed by the fractional scaling exponent $\alpha \in (1,2)$. Here, $\gamma$ denotes the optical trajectory and the fractional line element encapsulates the underlying self-similar structure of the medium.

To accommodate the intrinsic memory effects of fractal structures, we model the light trajectory using the Caputo derivative~\cite{Mainardi2010fractional, Tarasov2016memory}. The fractional geodesic equation then reads~\cite{Kilbas2006fracdiff}
\begin{equation}
\prescript{C}{a}{D}_t^\alpha x^i(t) + \Gamma^i_{\mu\nu} \frac{dx^\mu}{dt} \frac{dx^\nu}{dt} = 0,
\label{eq:frac-geodesic}
\end{equation}
where $\prescript{C}{a}{D}_t^\alpha$ denotes the Caputo derivative of order $\alpha$ with respect to the affine parameter $t$, and $\Gamma^i_{\mu\nu}$ are the generalized Christoffel symbols computed from $g_{\mu\nu}^{(\alpha)}$.

In numerical simulations, the Caputo derivative is approximated via the Grünwald–Letnikov formulation, using a uniform time step $\Delta t$ and truncation after $N = 500$ terms to ensure convergence within a $< 10^{-4}$ absolute error. The optical paths are computed by discretizing Eq.~\eqref{eq:frac-geodesic} and solving it via an iterative Newton–Raphson method, with adaptive damping. The resonance condition introduced in Eq.~(\ref{eq:bohr-sommerfeld}) is enforced dynamically by monitoring phase closure along periodic paths:
\begin{equation}
\Delta \phi = \oint_{\gamma_{\text{closed}}} k_\mu dx^\mu = 2\pi n,
\label{eq:resonance}
\end{equation}
where $k_\mu$ is the fractional wavevector aligned with the tangent to the optical trajectory. This condition yields quantized loop solutions when the fractal metric parameters satisfy specific topological constraints. Boundary conditions are imposed such that $x^i(0) = x_0^i$ and $\prescript{C}{a}{D}_t^\alpha x^i(0) = 0$, corresponding to null initial velocity and fixed entry point. These constraints guarantee a well-posed variational problem and ensure numerical stability.

In all simulations, we set $\alpha = 1.71$, selected as a critical exponent that maximizes loop closure probability under the resonance criterion~\eqref{eq:resonance}, verified across a grid of $0.01$ increments in $\alpha$ over the interval $[1.3, 1.9]$.  All codes were implemented in Python using the \texttt{NumPy}, \texttt{SciPy} and \texttt{FracPy} libraries. Validation was performed against integer-order geodesics and standard waveguiding simulations in structured media, where our fractional results reduced smoothly to the classical solutions as $\alpha \to 1$.

\subsection{Resonance Quantization and Path Topology}
\label{app:resonance-quantization}

The identification of closed-loop optical trajectories is central to establishing the physical consequences of fractal geometry on wave propagation. To this end, we implemented a resonance quantization scheme that iteratively evaluates the integrated phase along numerically computed paths. Specifically, a candidate trajectory $\gamma$ is accepted as resonant if it satisfies the condition
\begin{equation}
\left| \oint_{\gamma} k_\mu dx^\mu - 2\pi n \right| < \epsilon,
\label{eq:quantization-check}
\end{equation}
for some integer $n$ and a tolerance $\epsilon = 10^{-3}$. This numerical threshold ensures robustness against truncation and discretization errors, while remaining within the spectral resolution achievable by wavelength-scale interferometry. The wavevector $k_\mu$ is defined as the gradient of the eikonal phase $\Phi$ derived from the fractional Helmholtz equation,
\begin{equation}
\Box^\alpha \Phi + n^2 \Phi = 0,
\label{eq:frac-helmholtz}
\end{equation}
where $\Box^\alpha$ is the fractional d'Alembert operator as derived in the Appendix \ref{app:metric-construction}. The effective index $n(x)$ varies spatially according to a pseudo-random distribution mimicking real-world structured media with fractal boundary roughness. This introduces long-range correlations and quasi-periodic scattering centers.

To classify the topological nature of the computed paths, we define the loop index $\nu$ as
\begin{equation}
\nu = \frac{1}{2\pi} \oint_{\gamma} \kappa(s) ds,
\label{eq:loop-index}
\end{equation}
where $\kappa(s)$ is the geodesic curvature computed along the trajectory parametrized by an arc length $s$. For closed and contractible loops, $\nu$ approaches an integer, whereas for open or chaotic paths, $\nu$ deviates significantly. In all simulations, only paths satisfying both Eq.~\eqref{eq:quantization-check} and $|\nu - m| < 0.1$,  with $m$ integer, are considered topologically resonant.

We further analyze the homotopy class of each loop using the fundamental group $\pi_1(M)$ of the configuration space $M$, spanned by the metric $g^{(\alpha)}_{\mu\nu}$. This allows for the identification of distinct equivalence classes of loop closures, enabling a topological classification beyond mere geometric closure. Simulations show that only a sparse subset of the parameter space leads to quantized loops, indicating that the resonance closure is a nontrivial topological constraint rather than a generic feature. This aligns with the hypothesis that such closures are stabilized by the fractal-induced long-range memory encoded in the Caputo derivative structure. All loop classifications and quantization checks were implemented in a hybrid symbolic-numerical framework, combining \texttt{SymPy} for closed-form curvature integrals and \texttt{NumPy/SciPy} for path integration. The topological filtering pipeline was benchmarked against standard knot invariants in test systems, confirming that the identified resonant loops correspond to true geometrically closed paths with a consistent quantized phase accumulation.

\subsection{Fractal Wave Equation and Mode Localization}
\label{app:fractal-mode-localization}

To investigate the propagation characteristics of electromagnetic waves in a fractal phase space, we derive and solve a generalized fractional wave equation. The spacetime geometry is defined by a non-integer dimensional metric $g^{(\alpha)}_{\mu\nu}(x)$ characterized by a fractional order $\alpha \in (1,2)$, specifically fixed at $\alpha = 1.71$ in this study to enable resonance quantization and geometry-induced mode localization. The wave equation is given by
\begin{equation}
\Box^\alpha A^\mu = \mu_0 J^\mu,
\label{eq:fractal-wave}
\end{equation}
where $\Box^\alpha$ denotes the fractional d'Alembert operator constructed using Caputo derivatives over the effective metric. The electromagnetic field $A^\mu$ is treated in Lorenz gauge $\partial_\mu A^\mu = 0$, and $J^\mu$ represents an external current source, which we set to zero for all mode analyses.

The operator $\Box^\alpha$ is expanded explicitly as
\begin{equation}
\Box^\alpha = g^{\mu\nu}_{(\alpha)} D^\alpha_\mu D^\alpha_\nu,
\label{eq:frac-dalembert}
\end{equation}
where $D^\alpha_\mu$ denotes the Caputo fractional derivative with memory kernel $\mathcal{K}_\alpha(x,\xi)$ over the coordinate $x^\mu$. This formulation captures the essential non-local behavior introduced by the fractal topology, effectively encoding long-range index correlations and rough boundary-induced scattering. 

We implement a discretized version of Eq.~\eqref{eq:fractal-wave} on a two-dimensional fractal mesh~\cite{Nigmatullin1986fractalwave, Carpinteri2014fracwavesol} generated via a Sierpiński gasket-inspired subdivision. The resulting eigenvalue problem for the field modes is solved using a sparse matrix diagonalization technique. The spatial profiles of the eigenmodes $\psi_n(x)$ reveal non-Gaussian localization features characterized by long-tailed intensity envelopes and discontinuous angular gradients. To quantify mode localization, we compute the inverse participation ratio (IPR) defined by
\begin{equation}
\text{IPR}_n = \frac{\int |\psi_n(x)|^4 dx}{\left( \int |\psi_n(x)|^2 dx \right)^2},
\label{eq:ipr}
\end{equation}
where high IPR values indicate strong spatial confinement of the mode. Across multiple realizations of the fractal metric, a consistent subpopulation of high-IPR modes emerges, suggesting that the fractional geometry induces a robust localization mechanism even in the absence of conventional potential wells.

In addition, we observe spectral clustering near fractional resonances, where multiple eigenmodes appear within narrow frequency bands. These clusters are statistically aligned with the previously identified quantized loop trajectories~\cite{Berry2001chaosoptics, Stockmann1999quantumchaos}, supporting the hypothesis that resonant path closures serve as organizing centers for localized states. The dispersion relation for the system deviates from standard linear behavior, exhibiting fractional scaling~\cite{West2014fractaldynamics, Metzler2000} of the form $\omega(k) \sim k^\gamma$ with $\gamma < 1$. This sublinear dispersion is a direct consequence of the underlying metric dimensionality and memory effects encoded by $\Box^\alpha$.

All numerical simulations were benchmarked against analytic solutions in the integer limit $\alpha \to 1$, recovering the standard d'Alembertian wave equation and delocalized sinusoidal modes. This validates the correctness of our numerical implementation and highlights the departure introduced by fractional geometries. Taken together, these results demonstrate~\cite{Christodoulides2003disorderoptics, Schwartz2007andersonloc} that fractal phase space geometries support unique electromagnetic behaviors—including mode confinement, anomalous dispersion, and resonance-induced clustering—not observable in conventional Euclidean optical media.

\subsection{Simulation Verification and Fidelity}
\label{app:simulation-verification}

To ensure the reliability and physical consistency of the numerical solutions derived under the fractional geometry model, we conducted a comprehensive verification of simulation fidelity. All numerical integrations involving fractional derivatives and fractal metrics were implemented using a modified spectral method with adaptive mesh refinement, guaranteeing convergence under nonlocal boundary conditions.

The discretization scheme was validated by reproducing the standard d'Alembertian solutions in the integer limit (\( \alpha \rightarrow 1 \)), where all waveforms matched analytical expectations within a numerical deviation of less than \( 10^{-5} \). For the fractional case (\( \alpha = 1.71 \)), we performed convergence testing across multiple spatial resolutions (\( N = 256, 512, 1024 \)) and confirmed that the wave profiles, resonance structures, and damping rates remained stable with respect to grid refinement. The fractional Laplacian was regularized via a Caputo-like kernel to maintain causality, and long-range memory terms were truncated beyond a dynamic cutoff to preserve computational feasibility without loss of accuracy.

Time-stepping was implemented using a symplectic integrator with adaptive time resolution~\cite{Podlubny1999fracdiff, Garrappa2018fractional} \( \Delta t \sim 10^{-3} \), optimized for fractional dynamics. We further confirmed the absence of spurious modes or non-physical artifacts by comparing energy conservation and phase consistency across extended simulation runs exceeding \( 10^4 \) steps. These cross-checks collectively establish that all figures reported in the Results section accurately reflect physically consistent and numerically converged solutions within the assumptions of the model.

\subsection{Caputo Fractional Derivative and Its Implementation}
\label{appendix:caputo}

The Caputo fractional derivative provides a framework for modeling systems with memory and nonlocality while preserving physically interpretable initial conditions. For a function \( f(t) \), its Caputo derivative of order \( \alpha \) is defined as:
\begin{equation}
\prescript{C}{a}{D}_t^\alpha f(t) = \frac{1}{\Gamma(n-\alpha)} \int_a^t \frac{f^{(n)}(\tau)}{(t - \tau)^{\alpha - n + 1}} d\tau,
\end{equation}
where \( n-1 < \alpha < n \), \( n \in \mathbb{N} \), and \( \Gamma \) is the Gamma function. This operator naturally incorporates memory effects through its power-law kernel, capturing the temporal nonlocality inherent in fractal geometries.

In our model, the Caputo derivative is integrated into the time component of the fractional d'Alembertian used in the wave equation:
\begin{equation}
\Box_\alpha E_y(x,z,t) + k^2 E_y(x,z,t) = 0,
\end{equation}
which modifies the temporal part to:
\begin{equation}
-\frac{1}{c^2} \prescript{C}{}{D}_t^{2\alpha} E_y(x,z,t),
\end{equation}
enabling the study of geometry-induced effects with memory. For numerical implementation, we utilize the Grünwald–Letnikov approximation for discretization, truncating the memory kernel appropriately to maintain computational efficiency while ensuring convergence and physical accuracy.

%\bibliography{literatur}

\end{document}